\documentstyle[12pt]{article}

\begin{document}
\begin{titlepage}
\title{\begin{flushright} {\normalsize UUITP-- 27/1993 } \\
\end{flushright} \bigskip \bigskip
{\bf Regular Representation \\ of the
     Quantum Heisenberg Double \\ \{ $U_q(sl(2))$, $Fun_{q}(SL(2))$
\}
\\
     ( $q$ is a root of unity). }}
\author{\\ {\bf D. V. Gluschenkov}
\thanks{Supported
in part by a Soros Foundation Grant awarded by
the American Physical Society.} \thanks{On leave of absence from
LOMI, Fontanka 27, St.Petersburg, Russia.
$\;\;\;\;\;\;\;\;\;\;\;\;\;\;\;\;\;\;\;\;\;\;\;\;\;\;\;\;\;\;$
e-mail:
snake@diada.spb.su $\;\;\;\;\;$ and $\;\;\;\;\;$ snake@lomi.sbp.su}\\
{\bf A. V. Lyakhovskaya}
\thanks{On leave of absence from
LOMI, Fontanka 27, St.Petersburg, Russia.
$\;\;\;\;\;\;\;\;\;\;\;\;\;\;\;\;\;\;\;\;\;\;\;\;\;\;\;\;\;\;$
e-mail:
anechka@diada.spb.su $\;\;\;\;\;$ and $\;\;\;\;\;$
anechka@lomi.sbp.su}\\\\
Institute of Theoretical Physics, Uppsala University, \\
 Box 803 S-75108, Uppsala, Sweden.}
\date{October 1993}
\maketitle \thispagestyle{empty}
\begin{abstract}
     Pairing between the universal enveloping algebra $U_q(sl(2))$
     and the algebra of functions over $SL_q(2)$ is obtained in
explicit
     terms.
     The regular representation of the quantum double is constructed
     and investigated. The structure of the root subspaces of the
Casimir
     operator is revealed and described in terms of $SL_q(2)$
elements.
\end{abstract}
\end{titlepage}

\normalsize

\newtheorem{prop}{Proposition}
\newenvironment{eqa}[2]{\begin{equation} \label{#2}
\begin{array}{#1}}{\end{array} \end{equation}}

\section*{Introduction.}

  Introduction containes the self-consistent definition of the
 problem solved in this paper.

     The quantum double is a pair of dual Hopf algebras \cite{Dri}.

    Hopf algebra is an assosiative algebra $A$, equipped by the
     following mappings \cite{Dri}:

\begin{itemize}
     \item{ homomorphism $\Delta$, called comultiplication
             $ \Delta : A \rightarrow A \otimes A $, such that
  \[ ( {\rm id} \otimes \Delta ) \Delta = ( \Delta \otimes {\rm id} )
\Delta
  \]
          }
     \item{ homomorphism $\varepsilon$, called co-unit\footnote{
            For completness we might add homomorphism $\eta$, called
unity,
          $ \eta : {\bf C} \rightarrow A $ such that
          $ \eta (\lambda) = \lambda \cdot {\bf 1}$ where $\lambda$
is
              a complex number and {\bf 1 } is the unity of the
              algebra $A$.}
             $ \varepsilon : A \rightarrow A $, such that
  \[ (\varepsilon \otimes {\rm id}) \Delta =
     ({\rm id} \otimes \varepsilon ) \Delta = {\rm id} \]
          }
     \item{ antihomomorphism $S$, called antipode
             $ S : A \rightarrow A $ such that
     \[ m (S \otimes {\rm id}) \Delta = m ({\rm id} \otimes S) \Delta
=
     \eta \circ \varepsilon, \]
  where $m : A \otimes A \rightarrow A$ is the associative
multiplication.
          }
\end{itemize}

     The quantum deformation of the Lie algebra $sl(2)$
     ( $ [ H, X_{\pm}] = \pm 2 X_{\pm}, [ X_+, X_- ] = H $)
      is an example of a Hopf algebra.

       Let $q$ be a complex number, $|q| = 1$ and denote by $K$ the
       following exponential expansion\footnote{ it is more usual to
denote
       $ K = q^{H} $, see for example papers \cite{Kac},
       \cite{Lus1}, \cite{Lus2}, devoted to the
     representation theory of quantum groups at roots of unity
          but we want $K$ (not $K^{1/2}$) to appear in the
          comultiplication (\ref{DeltaKX}).  }
\begin{eqa}{c}{ourK}
            K = q^{\frac{H}{2}}, \\
\end{eqa}
       then the Hopf algebra attributes look like

\begin{eqa}{c}{KX}
   K X_{\pm} = q^{\pm 1} X_{\pm} K, \\
  X_+ X_- - X_- X_+ = \frac{K^2 - K^{-2}}{q - q^{-1}}; \\
                                                    \\
\end{eqa}
\begin{eqa}{c}{DeltaKX}
   \Delta K = K \otimes K, \\
   \Delta X_{\pm} = X_{\pm} \otimes K^{-1} + K \otimes X_{\pm}; \\
\end{eqa}
\begin{eqa}{c}{SeKX}
                                                             \\
   S(K) = K^{-1}, \;\; S(X_{\pm}) = - K^{-1} X_{\pm} K; \\
   \varepsilon (1) = 1, \;\; \varepsilon (K) = 1, \;\; \varepsilon
(X_{\pm}) = 0.
\end{eqa}

   This Hopf algebra is the "quantum universal enveloping algebra"
   $U_q(sl(2))$ \cite{Kac}, \cite{Lus2}.

 Consider the matrix elements $ a, b, c, d $ of $ g \in SL(2) $
\[ g = \left( \begin{array}{cc} a & b \\
                                      c & d
                \end{array} \right).  \]

 They are simplest functions on the group .

 The elements of the quantized algebra of functions
 are linear combinations of monomials $ a^{i}b^{j}c^{k}d^{l} $.
 The Hopf algebra structure of $SL_q(2)$ can be defined as follows
 \cite{FaReTa}, \cite{Soi}:

    Multiplication:
\begin{eqa}{cc}{abcd}
   ad - qbc = 1.  & \\
   ab = q ba & ac = q ca \\
   bd = q db & cd = q dc \\
   bc = cb, & ad - da = (q -q^{-1}) bc
\end{eqa}

    Comultiplication:
\begin{eqa}{ll}{Delta_abcd}
   \Delta a = a \otimes a + b \otimes c &
                   \Delta c = c \otimes a + d \otimes c \\
   \Delta b = a \otimes b + b \otimes d &
                   \Delta d = c \otimes b + d \otimes d

\end{eqa} 

    Antipode:
\begin{eqa}{ll}{S_abcd}
   S(a) = d, & S(b) = - q^{-1} b, \\
   S(c) = - q c, & S(d) = a.

                                                       \\
\end{eqa} 

   Co-unit\footnote{
Multiplication can be written in a compact form: $
\begin{array}{ccc}
 R g^1 g^2 = g^2 g^1 R, & where &
 R(q) = \left( \begin{array}{cccc}
          q^{-1/2} & 0 & 0 & 0 \\
            0 & q^{1/2} & q^{-1/2}-q^{3/2} & 0 \\
            0 & 0 & q^{1/2} & 0 \\
            0 & 0 & 0 & q^{-1/2}
          \end{array} \right),

\end{array} $
while comultiplication, antipode and co-unit look like
$ \begin{array}{ccc}
   \Delta g = g^1 \otimes g^2, & S(g) = g^{-1}, & \varepsilon(g) =
{\bf 1}.
\end{array} $
}:

\begin{eqa}{ll}{epsilon_abcd}
  \varepsilon (a) = 1, & \varepsilon (b) = 0, \\
    \varepsilon (c) = 0, & \varepsilon (d) = 1.

                                                       \\
\end{eqa} 


Now we have two infinite dimentional Hopf algebras
$ U_q(sl(2))$ and \\ $SL_q(2)$.
In order to obtain a double,
one have to construct a bilinear pairing
\begin{eqa}{c}{pairing}
 U_q(sl(2)) \otimes SL_q(2) \rightarrow {\bf C},
\end{eqa} 
such that
\begin{eqa}{c}{pairing1}
 \langle U, f_1 \cdot f_2 \rangle =
 \langle \Delta U, f_1 \otimes f_2 \rangle
\end{eqa} 
and
\begin{eqa}{c}{pairing2}
   \langle U_1 \cdot U_2, f \rangle =
 \langle U_1 \otimes U_2, \Delta f \rangle
\end{eqa} 
for any $U, U_1, U_2 \in U_q(sl(2))$ and $f, f_1, f_2 \in SL_q(2)$,
where
\[ \langle U_1 \otimes U_2, f_1 \otimes f_2 \rangle \equiv
 \langle U_1, f_1 \rangle \cdot \langle U_2, f_2 \rangle.
\]

So, in quantum double, the elements of $A^{\ast} = SL_q(2)$ are
linear
functionals on $A= U_q(sl(2)) $ and vice-verse.
In the next section we construct the pairing between $U_q(sl(2))$
and $Fun_{q}{SL(2)}$ and
reveal that the majority of functionals in both spaces are
linearly dependent or equal to zero. Factorizing them out we obtain
the
Quantum Heisenberg double which turns out to be finite-dimentional.

\section{Pairing}

 We assume that
\begin{eqa}{lll}{pairing_assume}
 \langle K, a \rangle = \xi, & \langle K, b \rangle = 0, &
 \langle K, c \rangle = 0, \\
 \langle X_+, b \rangle = \zeta, & \langle X_+, a \rangle = 0, &
 \langle X_+, c \rangle = 0, \\
 \langle X_-, c \rangle = \vartheta, & \langle X_-, a \rangle = 0, &
 \langle X_-, b \rangle = 0.
 \end{eqa}

\begin{prop}
 Supposing that relations (\ref{pairing1}, \ref{pairing2})
 and (\ref{pairing_assume}) are valid,
 we derive the formula\footnote{ $ [n]! \equiv
\frac{(q^n-q^{-n})(q^{n-1}-q^{-(n-1)}) \cdots (q-q^{-1})}{
(q-q^{-1})^n }. $}

\begin{eqa}{c}{explicit_pairing}
\langle X_+^k K^t X_-^i, a^n b^l c^m \rangle = \delta^{kl}
\delta^{im}
q^{\frac{t(n-k-i)+n(k+i)}{2}} [i]! [k]! \zeta^k \vartheta^i.
\end{eqa}
and conclude that in (\ref{pairing_assume})
\begin{itemize}
   \item{ $ \zeta$ and $\vartheta$ can be chosen arbitrarily
   (we fix $\zeta =1$, $\vartheta =1$).  }
   \item{ but $\xi$ must
          be set equal to $q^{1/2}$.\footnote{Note that
            if one uses $q_U$ in the
   relations of $U_q(sl(2))$ and $q_f$ in those of $SL_q(2)$, he
   will inevitably
   come to the equations $\langle K, a \rangle = q_U^{\frac{1}{2}}$
and
   $\langle K, a \rangle = q_f^{\frac{1}{2}}$, so that $q_U$ and
$q_f$
   must be taken equal.  } } 
\end{itemize}
\end{prop}
     The proof is based on inductive method and involves the formulae
     (\ref{pairing1}, \ref{pairing2}) in turn.

Paring (\ref{explicit_pairing}) for $q^N=1$ obeys the
periodicity conditions:
\begin{eqa}{c}{+2N}
 \langle X_+^k K^{t+2N} X_-^i, f \rangle = \langle X_+^k K^t X_-^i, f
\rangle \\
 \langle U, a^{n+2N} b^k c^i \rangle = \langle U, a^n b^k c^i \rangle
\\
 {\rm for} \;\;{\rm any} \;\; f \in A^{\ast}= SL_q(2), \;\; U \in
A=U_q(sl(2)), \;\; \\
 {\rm and} \;\; {\rm for} \;\; {\rm any} \;\; k,i,t,l,m,n
\end{eqa}

In order to eliminate the identical functionals we factorize over the
relations
\begin{eqa}{c}{fact}
 K^{2N} = 1, a^{2N} = 1.
\end{eqa}
It is easy to check that this factorization does not violate the
structures (\ref{KX})-(\ref{epsilon_abcd}) of the Hopf algebras
$A$ and $A^{\ast}$.
Note, that
$ K^{N}, a^{N} $ belong to the centers of the corresponding algebras,
while we have to factorize over $ K^{2N}, a^{2N} $.

The cases of odd and even $N$ turn out to be different.

{\underline {\it For odd $N$} } we have
\begin{eqa}{l}{=0}
 \langle X_+^k K^t X_-^i, a^n b^l c^m \rangle = 0 \;
  {\rm for}\;\; k \geq N \;\; {\rm or} \;\; i \geq N,
  {\rm or} \;\; l \geq N, {\rm or} \;\; m \geq N,\\
  {\rm and} \;\; {\rm any} \;\; t,n,l,m
\end{eqa}
since
\[ [n]! = 0 \;\; {\rm for} \;\; n \geq N. \]

In order to exclude the zero functionals we factorize over the
following relations
\begin{eqa}{c}{fact=0}
  X^N_{\pm}= 0, \;\; b^N = c^N = 0.
\end{eqa}
 This factorization is compatible with the Hopf algebra requirements
 (\ref{KX}) -- (\ref{epsilon_abcd}).

The finite-dimensional dual bases can be introduced in $U_q(sl(2)$
and
$SL_q(2)$:
\begin{eqa}{c}{baso}
\langle
  \frac{1}{[k]! [i]!}q^{-\frac{t(k+i)}{2}} X_+^k K^t X_-^i , b^l
f_{n}(a) c^m
\rangle = \delta^{kl} \delta^{im} \delta^{t-k+i,n},
\end{eqa}
where
\begin{eqa}{ll}{ranges_odd}
 k,i,l,m = 1, \ldots N, & t,n = 1, \ldots, 2N, \\
\end{eqa} 
and we use the notation
\[ f_n (a) = \frac{1}{2N} \sum_{j=0}^{2N-1} q^{-\frac{nj}{2}} a^j.
\]
The dimensions of $A$ and $A^{\ast}$ are $ {\rm dim} \, A = {\rm dim}
\, A^{\ast} =2N^3$.

At the same time {\underline {\it for even $N$} } $ [ \frac{N}{2} ] !
=0$ and we obtain
\begin{eqa}{l}{=0e}
  \langle X_+^k K^t X_-^i, a^n b^l c^m \rangle = 0, \;\;
  k \geq \frac{N}{2} \;\; {\rm or} \;\; i
 \geq \frac{N}{2}, \;\; {\rm or} \;\; l \geq \frac{N}{2},
 \;\; {\rm or} \;\; m \geq \frac{N}{2},
\end{eqa}
Therefore we must factorize over
\begin{eqa}{c}{fact_e}
  X^{\frac{N}{2}}_{\pm}= 0, \;\; b^{\frac{N}{2}} = c^{\frac{N}{2}} =
0.
\end{eqa}
Although the elements $X^{\frac{N}{2}}_{\pm}$, $b^{\frac{N}{2}}$,
$c^{\frac{N}{2}}$ do not belong to the centers, the factorization is
still selfconsistent.\footnote{
 For example, $
 [ X_+^{\frac{N}{2}} , X_- ] = 0 \;\; for \; even \; N.  $
                       }

The formulae (\ref{baso}) for dual bases are still valid,
  but (\ref{ranges_odd}) must be substituted by
\begin{eqa}{ll}{ranges_even}
 k,i,l,m = 1, \ldots \frac{N}{2}, & t,n = 1, \ldots, 2N.  \\
\end{eqa} 
The dimensions of $A$ and $A^{\ast}$ are $ {\rm dim} \, A = {\rm dim}
\, A^{\ast}=
 2N \cdot (\frac{N}{2})^2 = \frac{N^3}{2}$.

\begin{prop}
The pairing (\ref{pairing}) satisfies the relations
(\ref{pairing1}) and (\ref{pairing2}).
\end{prop}
    The relation (\ref{pairing1}) can be proved by direct
calculation,
   and it takes slightly more time and accuracy in the case of
    (\ref{pairing2}).
\newpage


\section{ $A = U_q(sl(2))$ acts on itself.}

The elements of $A = U_q(sl(2))$ can act on $A$ by left and right
multiplication.

To describe the structure of this representation, we need
some extra objects (see \cite{ours} for details):
\begin{enumerate}
   \begin{item} The Casimir operator
     \begin{eqa}{c}{Casimir}
         C = X_+ X_- + \frac{q^{-1}K^2 + q K^{-2}}{(q-q^{-1})^2} =
            X_- X_+ + \frac{qK^2 + q^{-1} K^{-2}}{(q-q^{-1})^2}.
     \end{eqa} 
   \end{item}

   \begin{item} The representation $v_j$ of $U_q(sl(2))$: \\
       \begin{eqa}{c}{repUq}
               Ke_{j}^{m} = q^{m}e_{j}^{m} \; , \\ \\
               X_{\pm}e_{j}^{m} =
                   \sqrt{(j \mp m)_{q}(j \pm m+1)_{q}}e_{j}^{m \pm 1}
       \end{eqa} 
     with the basis $e_{j}^{m} \;\; (m=-j \; , -j+1 \; , \ldots ,
j)$\footnote{
      $ (x)_q = \frac{q^x - q^{-x}}{q - q^{-1}} $ }.

  {\underline {\it For odd $N$}} $ 0 \leq j \leq \frac{N-1}{2} $, \\
dim $v_j = 2j +1$;

  {\underline {\it for even $N$}} $ 0 \leq j \leq \frac{N/2-2}{2} $,
dim $v_j = 2j+1$.
   The representation $v_j$ is shown at Figure \ref{Vv}. Each point
denotes
   one vector of the basis; $X_+$ moves up and $X_-$ moves down.
   \end{item}

   \begin{item} The representation $V_j$ of $U_q(sl(2))$. \\
     The representation $V_j$ described by the same formulae
(\ref{repUq})
     corresponds to
     a negative $j$ (see Figure \ref{Vv}):

 {\underline {\it for odd $N$}} $ \;\; -\frac{N}{2} \leq j \leq
-\frac{1}{2} $,
       dim $V_j =N+2j+1$;

  {\underline {\it for even $N$}} $ -\frac{N}{4} \leq j \leq
-\frac{1}{2} $,
        dim $V_j = \frac{N}{2}+ 2j +1$ .  \\

\begin{figure} \caption{$v_j$ and $V_j$} \label{Vv}
\begin{center} \begin{tabular}{cc}
 \begin{picture}(150,200)(0,0)

 \put(70,140){$v_j$}

 \multiput(70,100)(0,-20){3}{\circle*{3}} \put(80,100){$e_j$}
 \put(70,20){\circle*{3}} \put(80,20){$e_{-j}$}
 \multiput(70,30)(0,10){3}{\circle*{1}}

\put(40,10){\vector(0,1){110}} \put(44,110){$ln K$}

 \end{picture} &
\begin{picture}(150,200)(0,0)

 \put(50,140){$V_j$}

 \multiput(50,100)(0,-20){3}{\circle*{3}} \put(60,100){$e_{-j-1}$}
 \put(50,20){\circle*{3}}
 \multiput(50,30)(0,10){3}{\circle*{1}}

 \put(60,12){$e_{-\frac{N}{2}-j}$ for even $N$}
 \put(60,24){$e_{-N-j}$ for odd $N$}

\put(20,10){\vector(0,1){110}} \put(24,110){$ln K$}

 \end{picture}
\end{tabular} \end{center}
\end{figure}
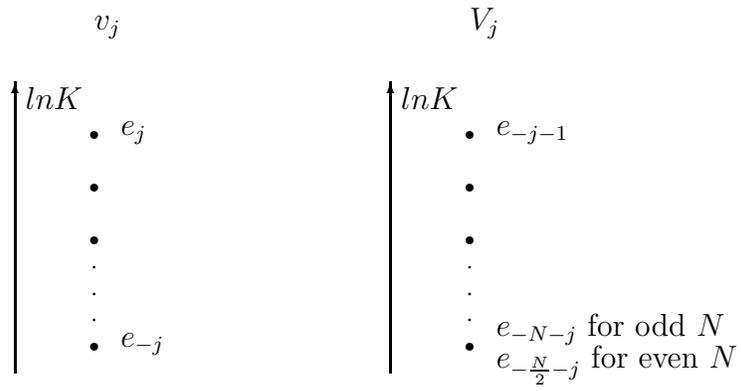

\begin{figure}
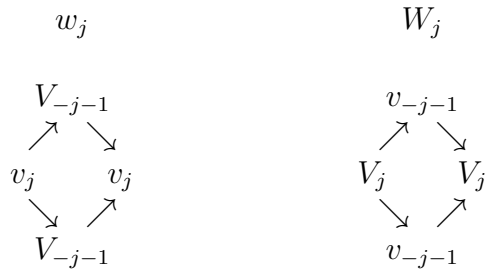
 \caption{$w_{j}$ and $W_j$} \label{Ww}
\begin{center} \begin{tabular}{ccc}
                  & & \\
                  & & \\
      $ \begin{array}{c}
                 w_j \\
                       \\
               V_{-j-1} \\
               \nearrow \;\;\; \searrow \\
               v_{j} \;\;\;\;\;\;\;\; v_{j} \\
               \searrow \;\;\; \nearrow \\
               V_{-j-1} \\
             \end{array} $ $\;\;\;\;$
            & $\;\;\;\;\;\;\;\;\;\;$ &
     $ \begin{array}{c}
                W_j \\
                     \\
               v_{-j-1} \\
               \nearrow \;\;\; \searrow \\
               V_{j} \;\;\;\;\;\;\;\; V_{j} \\
               \searrow \;\;\; \nearrow \\
               v_{-j-1} \\
             \end{array} $
\end{tabular} \end{center}
\end{figure}

       The two series $v_{j}$ and $V_{j}$
        form the complete list of irreducible representations
        of $U_q(sl(2))$.

   \end{item}

   \begin{item} The representation $w_j$.  \\
  {\underline {\it For odd $N$}} $ 0 \leq j \leq \frac{N}{2}-1 $, dim
$w_j = 2N$ ; \\
  {\underline {\it for even $N$}} $ 0 \leq j \leq \frac{N}{4}-1 $,
dim
$w_j = N$ .

    The structure of this representation is shown on Figure~\ref{Ww}.

    The following should be mentioned:

        a) Each letter denotes the array of vectors $e_m$ in the
        space of the corresponding representation: $e_m$
        under $e_{m+1}$ and so on.
        $X_+$ moves up and $X_-$ moves down. One might add a vertical
        coordinate axis on the figure and call it $ln K$.

        b) The meaning of the arrows: \\
\begin{tabular}{l}
                                      \\
        $X_{+}$ ({\rm the highest vector of the left copy of} $v_j$ )
= \\
         $\;\;\;\;\;\;\;\;\;\;\;\;\;\;\;\;\;\;\;$
                  ({\rm the lowest vector of the upper copy of} $V_j$
) \\

\\
        $X_{-}$ ({\rm the lowest vector of the upper copy of} $V_j$ )
= \\
            $\;\;\;\;\;\;\;\;\;\;\;\;\;\;\;\;\;\;\;$
               ({\rm the highest vector of the right copy of} $v_j$ )
\\

\\
         $X_{-}$ ({\rm the lowest vector of the left copy of} $v_j$ )
= \\
              $\;\;\;\;\;\;\;\;\;\;\;\;\;\;\;\;\;\;\;$
                 ({\rm the highest vector of the lower copy of} $V_j$
) \\

\\
        $X_{+}$ ({\rm the highest vector of the lower copy of} $V_j$
)
= \\
                $\;\;\;\;\;\;\;\;\;\;\;\;\;\;\;\;\;\;\;$
                  ({\rm the lowest vector of the right copy of} $v_j$
) \\
\end{tabular}

        c) Both $V_j$ and the right $v_j$
          are the eigenspaces of the Casimir operator with the
eigenvalue
            \begin{eqa}{l}{lamda}
       \lambda_j = \frac{q^{2j+1} + q^{-2j-1}}{(q+q^{-1})^2} =
               - \frac{\cos{\frac{2\pi}{N}(2j+1)}}{ 2
\sin^2{\frac{2\pi}{N}}}
            \end{eqa} 

{\it Each vector of the left copy of $v_j$ is the adjoint
          vector of the Casimir operator.
} The corresponding eigenvector
          is located in the right copy of $v_j$
          at the same "level" of $ln K$ scale.

          Thus, we have three types of vectors: adjoint vectors,
their
          eigenvectors and the eigenvectors which do not have adjoint
vectors.

   \end{item}

   \begin{item} The represetation $W_j$.  \\
  {\underline {\it For odd $N$}} $ \;\; -\frac{N}{2} \leq j \leq -1
$,
dim$W_j = 2N$ ; \\
  {\underline {\it for even $N$}} $ \;\;\;\;-\frac{N}{4} \leq j \leq
-1 $, dim$W_j = N$ .

    The structure of $W_j$ is shown on Figure~\ref{Ww}. Here again
    each vector of the left copy of $V_j$ is the adjoint
          vector of the Casimir operator.
\end{item}
\end{enumerate}


Now we concentrate on the structure of the regular representation
of $U_q(sl(2))$.

\newpage

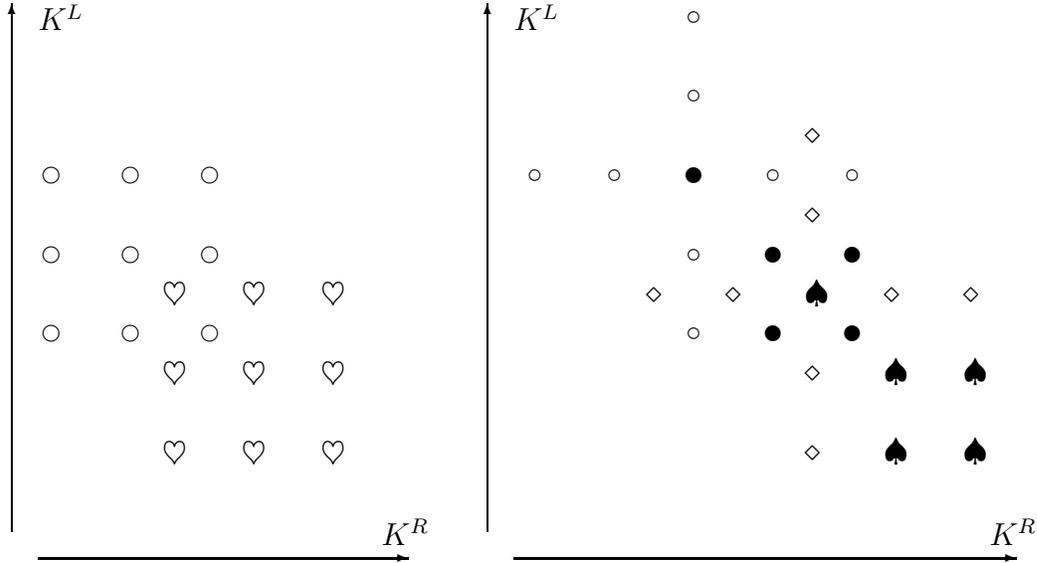
\begin{figure} \caption{Basis for $odd \;\; N$, $q^3=1$} \label{odd}
\begin{picture}(420, 250)(0,0)

\multiput(77,56)(30,0){3}{$\heartsuit$}
\multiput(35,105)(30,0){3}{\circle{6}}
\multiput(77,86)(30,0){3}{$\heartsuit$}
\multiput(35,135)(30,0){3}{\circle{6}}
\multiput(77,116)(30,0){3}{$\heartsuit$}
\multiput(35,165)(30,0){3}{\circle{6}}

\put(320,117){$\spadesuit$} \put(278,165){\circle*{6}}
\multiput(350,57)(30,0){2}{$\spadesuit$}
\multiput(308,105)(30,0){2}{\circle*{6}}
\multiput(350,87)(30,0){2}{$\spadesuit$}
\multiput(308,135)(30,0){2}{\circle*{6}}

\multiput(320,147)(0,30){2}{$\diamond$}
\multiput(278,195)(0,30){2}{\circle{4}}
\multiput(320,87)(0,-30){2}{$\diamond$}
\multiput(278,135)(0,-30){2}{\circle{4}}
\multiput(350,117)(30,0){2}{$\diamond$}
\multiput(308,165)(30,0){2}{\circle{4}}
\multiput(260,117)(30,0){2}{$\diamond$}
\multiput(218,165)(30,0){2}{\circle{4}}

\put(20,30){\vector(0,1){200}} \put(200,30){\vector(0,1){200}}
\put(30,20){\vector(1,0){140}} \put(210,20){\vector(1,0){190}}

\put(30,220){$K^L$} \put(210,220){$K^L$}
\put(160,25){$K^R$} \put(390,25){$K^R$}

\end{picture} \end{figure}

\begin{figure} \caption{Periodicity of the basis, $N=3$} \label{auxi}
\begin{picture}(420, 250)(0,0)

\put(35,165){$a$} \put(65,165){$b$} \put(95,165){$c$}
\put(35,135){$d$} \put(65,135){$e$} \put(95,135){$f$}
\put(35,105){$g$} \put(65,105){$h$} \put(95,105){$i$}

\put(80,120){$A$} \put(110,120){$B$} \put(140,120){$C$}
\put(80,90){$D$} \put(110,90){$E$} \put(140,90){$F$}
\put(80,60){$G$} \put(110,60){$H$} \put(140,60){$I$}

\put(77,57){\framebox(73,73){}}


\put(260,165){$a$} \put(290,165){$b$} \put(320,165){$c$}
\put(260,135){$d$} \put(290,135){$e$} \put(320,135){$f$}
\put(260,105){$g$} \put(290,105){$h$} \put(320,105){$i$}

\put(305,120){$A$} \put(335,120){$B$}
\put(305,90){$D$} \put(335,90){$E$}

\put(305,150){$G$} \put(335,150){$H$} \put(275,150){$I$}

\put(275,120){$C$} \put(275,90){$F$}

\put(302,89){\line(0,1){42}} \put(346,131){\line(-1,0){44}}
\put(302,147){\line(0,1){20}} \put(302,147){\line(1,0){44}}
\put(286,147){\line(0,1){20}} \put(286,147){\line(-1,0){20}}
\put(286,89){\line(0,1){42}} \put(286,131){\line(-1,0){20}}



\put(20,30){\vector(0,1){200}} \put(230,30){\vector(0,1){200}}
\put(30,20){\vector(1,0){140}} \put(240,20){\vector(1,0){160}}

\put(30,220){$K^L$} \put(240,220){$K^L$}
\put(160,25){$K^R$} \put(390,25){$K^R$}

\put(200,135){$\Longrightarrow$}

\end{picture} \end{figure}
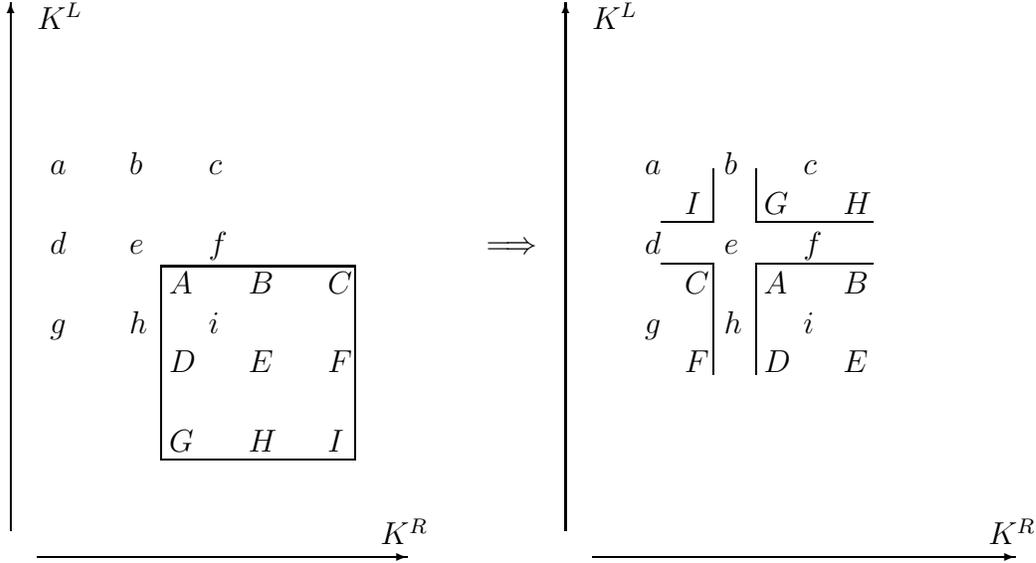

\newpage
The Casimir operator (\ref{Casimir})
decomposes the functional space into root subspaces of two different
types.

\begin{enumerate}
  \begin{item}
    { \bf Exceptional subspaces.}

        {\underline { \it For odd N }} there is one exceptional
subspace.
        It corresponds to $\lambda_{j = -\frac{1}{2}}$ in
(\ref{lamda}).
        For exceptional $j$, $j = -1/2$
         the representation $V_{-1/2} $ has the maximal possible
         dimension equal to $N$.

         This exceptional subspace consists of $2N^2$ Casimir
eigenvectors,
         which, at the same time, are eigenvectors for both left
         and right action of the operator $K$. These vectors may be
         represented as two $N \times N$ squares on the plane with
         coordinates $ln K^L$, $ln K^R$.

          Each column of circles
          at the diagram for $N=3$ (Figure \ref{odd}, left part)
          denotes the representation $V_{j = -1/2}$ of the left
$U_q(sl(2))$
          action.
          Each row of circles is the same representation
          of the right $U_q(sl(2))$ action.
          The other part of the subspace is represented by hearts.
          Each row of hearts is the representation $\tilde{V}_{j =
-1/2}$,
          which is described by the formulae:
       \begin{eqa}{c}{Vtilda}
               Ke_{j}^{m} = q^{m+N/2}e_{j}^{m} \; , \\ \\
               X_{\pm}e_{j}^{m} =
               - \sqrt{(j \mp m)_{q}(j \pm m+1)_{q}}e_{j}^{m \pm 1}.
       \end{eqa} 
          Thus the subspace
          consists of two parts which cannot be connected by
          $U_q(sl(2))$ action.

           One more comment must be add here.
           Due to the condition $K^{2N} = 1$ we have the torus
           ($ln K^L$, $ln K^R$) instead of the plane.
           Let us substitute dots by lowercase letters and hearts by
           capital letters. ( See the left side of Figure \ref{auxi}.
)
           If we want
           the torus to be presented as a square with the periodicity
           condition, we obtain the more precise
           ( but less transparent ) diagram at the right side of
           Figure \ref{auxi}. This periodicity must be taken into
account while
           examining more complicated diagrams (the right sides of
           Figures \ref{odd} and \ref{even}) but will not be
mentioned
           below.

    {\underline { \it For even $N$ }} there is one more exceptional
subspace.
           It corresponds to $\lambda_{\frac{N/2-1}{2}}$, where
         $N/2$ is the maximum dimension of $v_j$ for even $N$.
         The structure of both exceptional subspaces is the same and
it is presented
         at the left side of Figure \ref{even} for $N=6$.
         There are only two differencies between this scheme and the
         left side of Figure \ref{odd}.
         First, the size of each square now equals to $N/2$ (not to
$N$),
         though it still coincides with the maximum possible
dimension
         of $V_j$ and $v_j$. Second, the relative location of the
         squares is different.

  \end{item}

  \begin{item}
    { \bf Typical subspaces.}
      \begin{itemize}
         \begin{item}
    {\underline {\it For odd $N$}} there are $(N+1)/2$ different
eigenvalues
             (\ref{lamda}), one exceptional subspace and $(N-1)/2$
             typical subspaces.
         \end{item}
         \begin{item}
    {\underline {\it For even $N$}} there are $(N+2)/2$ different
eigenvalues
             (\ref{lamda}), two exceptional subspaces and $(N-2)/2$
             typical subspaces.
         \end{item}
      \end{itemize}

\newpage
\begin{figure} \caption{Basis for $even \;\; N$, $q^6=1$}
\label{even}
\begin{picture}(500,300)(0,0)

\multiput(119,57)(30,0){3}{$\heartsuit$}
\multiput(35,153)(30,0){3}{\circle{6}}
\multiput(119,87)(30,0){3}{$\heartsuit$}
\multiput(35,183)(30,0){3}{\circle{6}}
\multiput(119,117)(30,0){3}{$\heartsuit$}
\multiput(35,213)(30,0){3}{\circle{6}}

\put(379,118){$\spadesuit$} \put(290,213){\circle*{6}}
\multiput(409,58)(30,0){2}{$\spadesuit$}
\multiput(320,153)(30,0){2}{\circle*{6}}
\multiput(409,88)(30,0){2}{$\spadesuit$}
\multiput(320,183)(30,0){2}{\circle*{6}}

\multiput(380,149)(0,30){2}{$\circ$}
\multiput(290,243)(0,30){2}{\circle{4}}
\multiput(380,88)(0,-30){2}{$\circ$}
\multiput(290,183)(0,-30){2}{\circle{4}}
\multiput(410,118)(30,0){2}{$\circ$}
\multiput(320,213)(30,0){2}{\circle{4}}
\multiput(318,118)(30,0){2}{$\circ$}
\multiput(230,213)(30,0){2}{\circle{4}}

\put(20,30){\vector(0,1){260}} \put(220,30){\vector(0,1){260}}
\put(30,20){\vector(1,0){170}} \put(230,20){\vector(1,0){230}}

\put(30,280){$K^L$} \put(230,280){$K^L$}
\put(190,25){$K^R$} \put(450,25){$K^R$}

\end{picture} \end{figure}
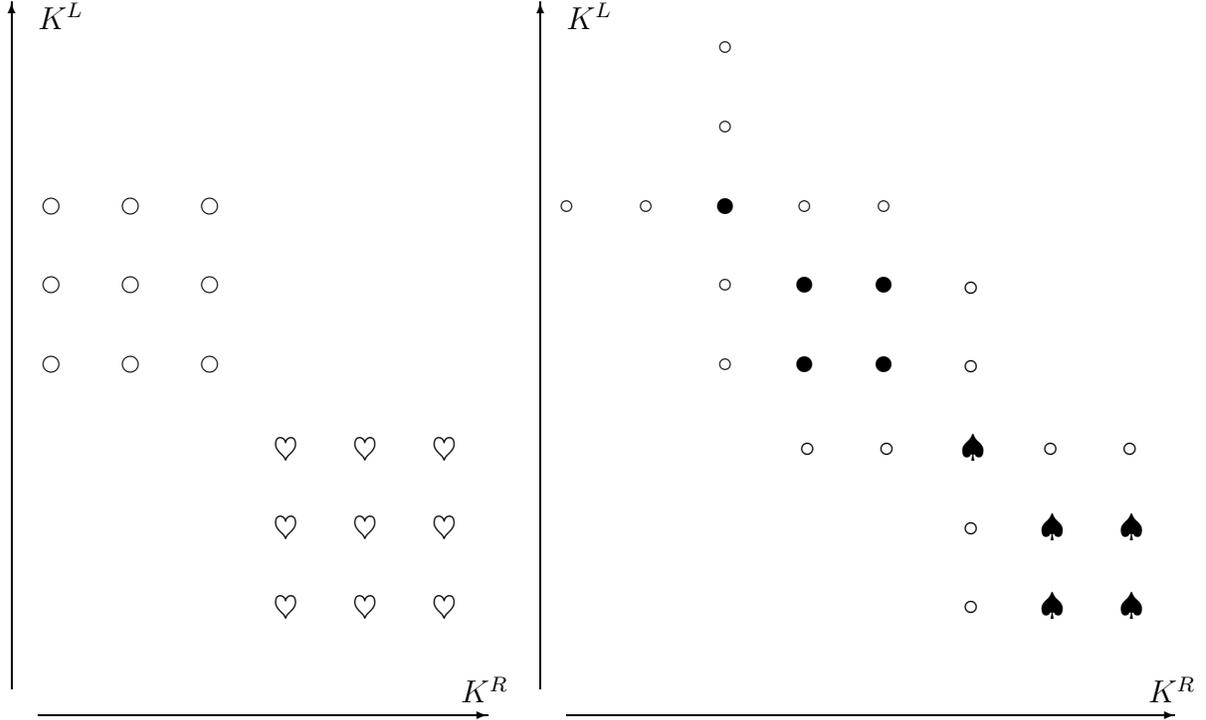

         The sum of dimensions of the partner representations
         $v_j$ and $V_{-j-1}$ in $w_j$ is always equal to the same
         number $M$ ($M = N$ if $N$ is odd, $M = N/2$ for even $N$)
         The same is valid for $V_j$ and $v_{-j-1}$
         in $W_j$.\footnote{
            Only $v_{\frac{M-1}{2}}$, ${\rm dim}v_{\frac{M-1}{2}} =
M$
            does not have a partner, since
            the dimension of this unexisting parnter would have been
equal
            to zero. Therefore $w_{\frac{M-1}{2}}$ does not exist.
            For the same reason $V_{-\frac{1}{2}}$,
            ${\rm dim}V_{-\frac{1}{2}} = M$
            does not have a partner, and there is no
$W_{-\frac{1}{2}}$.
            These exeptions correspond to exceptional subspaces.}
         We also have
\[
           \lambda_j = \lambda_{-j-1}.\footnote{
                 while for exceptional values of $j$ we obtain
simply:
                 $-j-1 = j$ $({\rm mod} N)$.}
\]
         Therefore, the eigenvalues of the Casimir operator coincide
         for $w_j$ and $W_{-j-1}$ ($ j \geq 0 $), and these
         representations are embedded into the same root subspace.

         The structure of the typical root subspaces is illustrated
         by the right sides of Figures \ref{odd} and \ref{even} for
odd and even $N$
         respectively. Each column corresponds to the representation
$w_j$ or
         $W_{-j-1}$ of the left action of
          $U_q(sl(2))$. Each row denotes the representation $w_j$ or
         $W_{-j-1}$ of the right action of $U_q(sl(2))$.
         Each white object represents one eigenvector, while each
         black object represents a pair: an adjont vector and
         its eigenvector. ( Remember the description of $w_j$
         and $W_j$).

         There is a principal difference between odd and even cases:
           \begin{itemize}
             \begin{item}
                 In the case of odd $N$ the subspace consists of
                 two parts which cannot be connected by
                 $U_q(sl(2))$ action: any vector in one part being
                 chosen as the initial will never be transfered to
the
                 second part with the help of $U_q(sl(2))$ action.
                 Starting from spades, you'll never get
                 to circles, neither white nor black ones.
             \end{item}
             \begin{item}
                 In the case of even $N$ there is no such
decomposition.
                 If you start from an adjoint vector in the spade
                 area, you easily get to some eigenvectors (not
                 adjoint ones!), located in black circles area -
                 via eigenvectors, represented by white circles.
             \end{item}
           \end{itemize}
  \end{item}
\end{enumerate}

The following rules describe the multiplication table for the
vectors of the space of $U_q(sl(2))$ representation. The zero
products are not mentioned.
\begin{enumerate}
  \begin{item}
     The eigenvalue of $K^R$ of the left multiple must coincide
     with the eigenvalue of $K^L$ of the right multiple in order to
     get a non-zero product. In terms of picture objects this means
     that the column containing the left multiple must intersect
     with the row containing the right multiple at the point where
     the eigenvalues of $K^R$ and $K^L$ coincide (i.e., on the common
     diagonal of the squares).
  \end{item}

  \begin{item}
    The eigenvalue of $K^R$ of the product equals to that of the
right
    multiple, and the eigenvalue of $K^L$ of the product equals
    to that of the left multiple. This means the product will be
placed
    at the junction of the left multiple row and the right multiple
    column.
  \end{item}

  \begin{item}
     \begin{tabular}{l}
    (adjoint vector) $\cdot$ (adjoint vector) = (adjoint vector) \\
    (eigenvector) $\cdot$ (adjoint vector) = (eigenvector) \\
    (eigenvector) $\cdot$ (eigenvector) = (eigenvector) \\
    (eigenvector which has an adjoint one) $\cdot$ (eigenvector) = 0
\\
     \end{tabular}
  \end{item}
\end{enumerate}

\section{ Regular representation of $U_q(sl(2))$.  }

We investigate the regular representation of $U_q(sl(2))$ on the
space
of functions over $SL_q(2)$.

The elements of $A = U_q(sl(2))$ can act
\begin{enumerate}
  {\item on $A$ by left and right multiplication, }
  {\item on $A^{\ast}$, according to the following definition:
\begin{eqa}{lr}{XonF1}
  U^L (f) = \langle U, \Delta f \rangle_1 & {\rm left} \;\; {\rm
action} \\
  U^R (f) = \langle U, \Delta f \rangle_2 & {\rm right} \;\; {\rm
action} \\
\end{eqa}
  }
\end{enumerate}

But if one identifies each element of $f_j \in A^{\ast}$ with the
corresponding
element $V_i \in A$: $ \langle V_i, f_k \rangle = \delta_{ik} $
then he finds that these two representations are conjugate:
\begin{eqa}{c}{conju}
    U^L(V_a) = V_b \;\;\;\;\; \Rightarrow \;\;\;\;\; U^L (f_b) = f_a
\\
    U^R(V_a) = V_b \;\;\;\;\; \Rightarrow \;\;\;\;\; U^R (f_b) = f_a
\\
\end{eqa} 

Therefore the space of the representation of $U_q(sl(2))$ can be
regarded,
first, as $A$, and second, as $A^{\ast}$. The action of
$A = U_q(sl(2))$ on $A$ is already described in the previous section.
The regular representation
of $A = U_q(sl(2))$ illustrated on Figures \ref{odd}, \ref{even}
acts on the elements of $SL_q(2)$ and
has evidently the same structure.
Now we investigate this representation in terms of
elements of $SL_q(2)$.

Applying the definition (\ref{XonF1}) and taking into account pairing
(\ref{explicit_pairing}) one easily obtains:
\begin{eqa}{cc}{K_on_a}
   X_-^L a = 0, \;\; K^L a = q^{1/2} a \\
\end{eqa} 
Acting by the Casimir operator (\ref{Casimir})
on monomials $a^k, k \leq 2N$, we see that these monomials are
eigenvectors with eigenvalues $\lambda_{j = \frac{k}{2}-1}$.
These eigenvalues coincide if
\begin{eqa}{lll}{k=k}
k = k^{\prime} \;\; ({\rm mod}N), & or & k + k^{\prime} = 2 \;\;
({\rm
mod}N). \\
\end{eqa}
For example, for $N=3$, $a^2$ belongs to the exceptional subspace
$j=-1/2$,
and $1=a^0$ and $a^1$ are contained in the typical subspace
$j=0,-1$, unique for the given $N$.

Figure \ref{q^3} represents the circles from the Figure \ref{odd}
(adjoint vectors are enclosed in square brackets.).
Consider the table $3 \times 3$.
Since the representation under discussion is conjugate to one
considered
in the previous section, $X^L_{-}$ now goes {\it up} and kills the
upper
row of the table. This follows from the relations
$X_-^L a = 0, X_-^L b = 0$. The $3 \times 3$ table
is a good illustration of the general situaion: for any $N$ and
any root subspace the upper row of such a square of eigenvectors
consists of $a^{m}b^{l}$, the left column consists of $a^{m}c^{l}$,
and
so on. Correspondence between the power of $a$ in the upper-left
corner and the eigenvalue was mentioned above. Starting from this
$a^k$
and moving it down by $X_+^L$ and right by $X_-^R$ one can derive
the whole square of eigenvectors.

Now consider the adjoint vector $q^2abc$ corresponding to the
eigenvector
$a$. Due to the relations
\[ K^{L}(f \cdot (bc)) = K^{L}(f), \;\;\;\; K^{R}(f \cdot (bc)) =
K^{R}(f),
\] the eigenvalues of $K^{L}, K^{R}$ remain unchangeable, but $c$
enables
$X_-^L$
to make {\it one and only one} step up ( $a$ and $b$ would be killed
)
and
to get to the eigenvector $q^2a^2b$ (next step up leads to zero).
At the same time the second power of $c$ in the adjoint vector
$b^2c^2 + bc$ enables $X_-^L$ to make {\it two} steps up.
This is the general rule:
an adjoint vector is simply equal to its eigenvector, multiplied by
a polynomial in $bc$
$(\alpha_0 + \alpha_1 bc + \cdots + \alpha_{s} (bc)^{s})$,
where the highest power $s$ is constant for the whole square with
the eigenvector $a^k$ in the upper-left corner, and the value of $s$
allows
to perform the required number of steps:
 $ s = N - k- 1 $ {\underline {\it for odd $N$}}; $ s = N/2 - k- 1 $
 {\underline {\it for even $N$}}.

Let's turn to the description of eigenvectors which do not have
adjoint ones.
There are two upper-right corners of "wings" on figure \ref{q^3}.
The two corner vectors must be killed by the action of corresponding
operators (moving them up or to the right), and consequently
each of them must look like
\begin{eqa}{rcl}{wing_odd}
 a^{\alpha} b^{N-1} & = & d^{2N - \alpha} b^{N-1} \\
\end{eqa}
for odd $N$, and
\begin{eqa}{rcl}{wing_even}
 a^{\alpha} b^{N/2-1} & = & d^{2N - \alpha} b^{N/2-1} \\
\end{eqa}
for even $N$.

 For example,
when going right $ a b^{N-1}$ is transfered to $ b^{N} = 0 $.
At the same time after some steps left we must obtain the eigenvector
$a^k$.
Or after some steps down we must obtain $d^k$.
This determines $\alpha$ from (\ref{wing_odd}, \ref{wing_even}):

\begin{eqa}{rcl}{pow_odd}
\alpha + (N-1) = k-1 \; ({\rm mod} 2N)
\end{eqa}
for odd $N$, and
\begin{eqa}{rcl}{pow_even}
\alpha + (N/2-1) = k-1 \; ({\rm mod} 2N)
\end{eqa}
for even $N$.
All the other vectors of upper-right wings can be obtained simply
using the formulae for $X_{\pm}^{L,R}$ action.
These vectors can also be calculated acting on the
adjoint vectors in the square,
but in this case the calculations are more complicated.

\begin{figure} \caption{Circles from Figure~3 in terms of $a,b,c,d$}
\label{q^3}
 \begin{tabular}{ccc}
          & & \\
          & & \\
  $\;\;\;a^2\;\;\;$ & $\;\;\;-qab\;\;\;$ & $\;\;\;-b^2\;\;\;$ \\
          & & \\
  $\;\;\;qac\;\;\;$ & $\;\;\;q^2 bc + ad\;\;\;$ & $\;\;\;-q^2
db\;\;\;$ \\
          & & \\
  $\;\;\;-c^2\;\;\;$ & $\;\;\;q^2 dc\;\;\;$ & $\;\;\;d^2\;\;\;$ \\
          & & \\
\end{tabular}

\begin{tabular}{ccccc}
           & & & & \\
           & & & & \\
           & & & & \\
           & & $\;\;\; -qdb^2\;\;\;$ & & \\
           & & & & \\
           & & $\;\;\; qd^2 b\;\;\;$ & & \\
           & & & & \\
 $\;\;\;-qdc^2 \;\;\;$ & $\;\;\;qd^2 c \;\;\;$ & $\;\;\; {\bf 1, [b^2
c^2 +bc]} \;\;\;$ & $\;\;\;q^2 a^2 b\;\;\;$ & $\;\;\; -q^2
ab^2\;\;\;$
\\
           & & & & \\
           & & $\;\;\; q^2 a^2 c\;\;\;$ & $\;\;\; {\bf a, [q^2 abc]}
\;\;\;$ & $\;\;\; {\bf b, [-b^2 c + (1- \alpha) b]} \;\;\;$ \\
           & & & & \\
           & & $\;\;\;-q^2 ac^2\;\;\;$ & $\;\;\; {\bf c, [-bc^2
+\alpha c]} \;\;\;$ & $\;\;\; {\bf d, [qdbc]} \;\;\;$ \\
           & & & & \\
 \end{tabular}

\end{figure}

The vectors already described in terms of $a, b, c, d$ (Figure
\ref{q^3})
are shown by black and white circles on Figure \ref{odd}.
 The second half of the vectors can be obtained multiplying
the expressoins from (Figure \ref{q^3}) by $a^N$ and
changing the signes of chess-like located
elements. This follows from the
relation
\[ X_{\pm} v = u \;\;\; \Rightarrow \;\;\; X_{\pm} a^N v =
q^{\frac{N}{2}} a^N u = - a^N u.  \]
The eigenvalues of the operator $K$ are multiplied by
$q^{\frac{N}{2}}$:
\[ Ku = \lambda u \Rightarrow K a^N u = q^{\frac{N}{2}} \lambda u. \]
Consequently, the second copies
must be shifted by $\frac{N}{2}$ in the scale of $ln K^L$ and $ln
K^R$.

Now we have the complete description of the regular representation of
$U_q(sl(2))$. Let us turn to the joint representation of
$U_q(sl(2))$ and $Fun_{q}(SL(2))$.

\section{Joint representation of $U_q(sl(2))$ and $Fun_{q}(SL(2))$.}

The elements of $A^{\ast} = SL_q(2)$ can act
\begin{enumerate}
  {\item on $A^{\ast}$ by left and right multiplication, }
  {\item on $A$, according to the following definition:
\begin{eqa}{lr}{XonF2}
  f^L (U) = \langle \Delta U, f \rangle_1 & {\rm left} \;\; {\rm
action} \\
  f^R (U) = \langle \Delta U, f \rangle_2 & {\rm right} \;\; {\rm
action} \\
\end{eqa}
  }
\end{enumerate}

The last proposition should complete the investigation of the regular
representation of the double.

\begin{prop} The joint regular representation of the double
$(U_q(sl(2)),SL_q(2))$ on the space of functions over $SL_q(2)$ is
irreducible.
\end{prop}

This statement can be proved in three steps:
1. Any polinomial of $a,b,c,d$ can be transfered to the combination
$\sum \gamma_k a^k$ by the action of suitable $X^{L,R}_{\pm}$.

2. The action of $(X^L_+)^l$ on this combination leads to $c^l$, $l$
is the
largest power in the sum $\sum \gamma_k a^k$. Then the action of
$X^L_-$ draws it back to $a^l$.

3. The monomial $a^l$ can be transmitted to the sector of given
Casimir
eigenvalue multiplying it by $a^i$, then, multipling $a^i$ by
$(\alpha_0 + \alpha_1 bc + \cdots $ \\ $ + \alpha_{s} (bc)^{s})$
$\in$
$SL_q(2)$,
one obtains the adjoint vector.
Finally, the aciton of $X_{\pm}^{L,R}$ can transmit it to any vector
in
the wings.

Thus, any function of $a,b,c,d$ can be obtained from any fixed
function
by the action of the regular representation. This proves the
Proposition.

\vspace*{3cm}

\section*{Acknowledgements.}

We are very grateful to L.D.Faddeev and M.A.Semenov-Tian-Shansky for
fruitful
advices and educative discussions.
We would like to thank Prof. A.Niemi
 for the hospitality and the excellent conditions
at the University of Uppsala
which made it possible to carry out this work. D.G. is grateful to
the Soros Foundation for the financial aid.
Special thanks to A.Alekseev for important comments and
careful reading of the preliminary version of the paper.


\vspace*{3cm}

\begin{thebibliography}{7}
\bibitem{Dri}{V.Drinfeld, Quantum groups.-{\it Proc. IMC-86} Berkeley
1986,
v.1}
\bibitem{ours}{A.Alekseev, D.Gluschenkov, A.Lyakhovskaya, Regular
representation of the quantum group $sl_q(2)$ ($q$ is a root of
unity).-
 {\it preprint LPTHE}, PAR-LPTHE 92-24,1992, to be published in
Algebra i Analiz.}
\bibitem{Kac}{V.Kac, C.De Concini, Representations of quantum groups
at
roots of 1.- {\it Colloque Dixmier}, 1990 ({\it Progress in
mathematics},
Birkh\"auser).}
\bibitem{Lus1}{G.Lusztig, Finite dimensional Hopf algebras arising
from
quantized universal enveloping algebras.- {\it J. Am. Math. Soc.},
v.3, N 1, 1990.}
\bibitem{Lus2}{G.Lusztig, Quantum groups at roots of 1.- {\it Geom.
Dedi.,} 35, 1990.}
\bibitem{FaReTa}{N.Reshetikhin, L.Takhtajan, L.Faddeev, Quantization
of Lie
grtoups and Lie algebras.- {\it Algebra i Analiz},1989, v.1 N 1
(Russian).}
\bibitem{Soi}{Ya.Soibelman, Algebra of functions on a compact quantum
group and its representations.- {\it Algebra i Analiz}, 1990, v.2 N 1
(Russian).}
\end{thebibliography}
\end{document}